\documentclass{elsart}

\usepackage{epsfig}
\usepackage{amssymb}

\begin{document}

\begin{frontmatter}

\title{Using Monte Carlo to optimize variable cuts}
\author{Erik Elfgren}
\ead{elf@ludd.ltu.se}

\begin{abstract}
A Monte Carlo method to optimize cuts on variables is presented
and evaluated. The method gives a much higher signal to noise
ratio than does a manual choice of cuts.
\end{abstract}
\end{frontmatter}

%\tableofcontents\pagebreak
%\section{Introduction}
There are two important methods for refining a signal over background ratio:
likelihood analysis and cut-based analysis. Likelihood analysis has the advantage
of not discarding any potential signal. However, it is not as straightforward
to evaluate its statistical significance compared to the cut-based analysis.
Cut-based analysis on the other hand, cuts away parts of the signal in
order to reduce the background evenmore.

%things that are known to
%be characteristic for the background, while leaving the signal as intact as
%possible, though inevitably reducing it.

Traditionally, variable-cuts have been sought with the help of good sense and
some experimentation.

In this letter I address an automatical method for
searching for optimal cuts. I have used only one simulated set of data, fairly
large, and many different backgrounds. The simulated data are heavy leptons
with masses $100-200$~GeV done at center-of-mass energies of $183-209$~GeV to
correspond to the OPAL experiment at LEP.

%\section{Method}
The method is simple:
Initially, determine which of the variables are most relevant and what their
ranges are. If possible, find some minimum cuts that will leave the
signal intact, while still reducing the background. This can significantly reduce 
the time spent on each iteration below.

The cut optimization then has the following general algorithm:
\begin{enumerate}
\item Choose a random variable and change the cut randomly with a value between
	0 and $T*max$, where $T$ is initialized as $T_i=100$\% and $max$ as the
	maximum value of the variable.
\item If this change leaves us with a higher $S/\sqrt{B}$-value, keep it,
	otherwise discard it.
\item Decrease $T$ and restart from the beginning
\end{enumerate}
A problem with this method is that it might get stuck in a local minimum somewhere.
This can be remedied by storing the final cuts and the $S/\sqrt{B}$-value and
then reinitializing the process, iterating until a satisfying $S/\sqrt{B}$-value
is obtained.
The method can be parametrized by $\Delta T$, the change in $T$ per iteration,
$T_{i}$, the initial value of $T$ and $N_{it}$, the number of reinitializations.

%\section{Test case}
Our test case is described in general in \cite{2000EPJC...14...73O}
and in particular in \cite{Elfgren:2002ck}. A short resum{\'e} follows here.
The signal we are looking for is $e^+e^-\rightarrow \bar\nu N \rightarrow \nu l qq$
and the main variables are the lepton energy $E_l$, the missing energy, $E_\nu$,
the invariant mass of $l$ and $\nu$, the invariant mass of the $N$ ($ =q,q,l$)
and the lepton type ($l=e,\mu$ or $\tau$).
Both the signal events and the background
events were subject to the full OPAL detector simulation \cite{1992NIMPA.317...47A}
as well as some basic cuts to ensure a good quality \cite{Alexander:1991qw}.
The miminum cuts mentioned above were set to $E_l, E_\nu \ge 5$ GeV.
The Monte Carlo generator EXOTIC \cite{EXOTIC} was used to generate the $e^+e^-\rightarrow \bar \nu N$
signal. The following masses were simulated $M_N=100, 110, 120, 130, 140, 150, 160,
170, 180, 190, 200$ GeV and for each mass the energies $E=183, 189, 192, 196, 200, 202, 204,
205,206,207,208$ GeV for all $M_N<E$.
The total number of signal events surviving the initial cuts were about 350 for each pair of ($E, M_N$).
A variety of MC generators was used to study the multihadronic background from SM,
see \cite{Elfgren:2002ck} and references therein. The relevant backgrounds are $qq\gamma$ (KK2f+PYTHIA 6.125), $llqq$, 
$eeqq$, $qqqq$, $ee\tau\tau$ (grc4f 2.1) and $\gamma\gamma q q$ (HERWIG).

\begin{figure}
\center
        \resizebox{0.8\hsize}{!}{\includegraphics{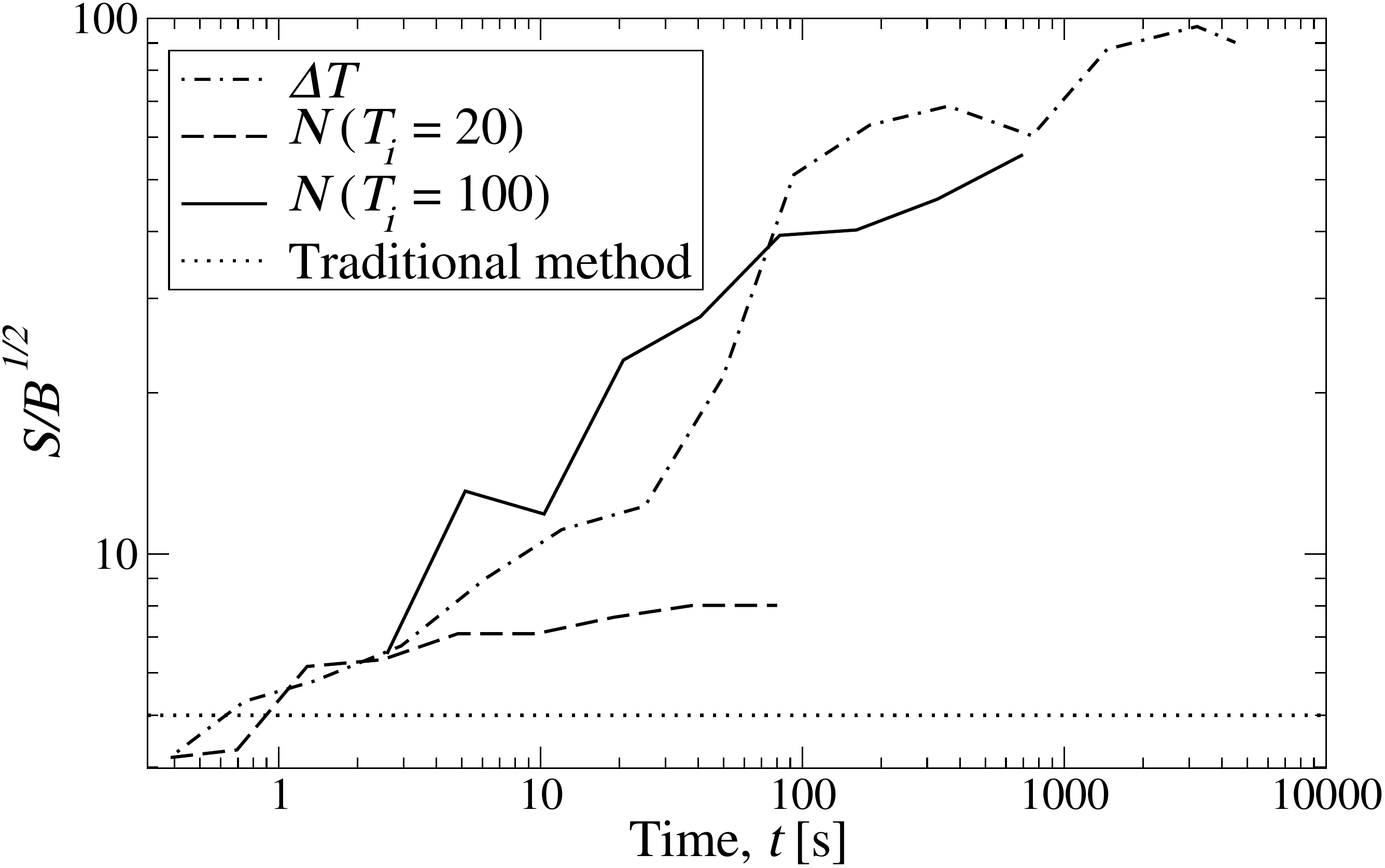}}
        \caption{Comparison of changes in some parameters.
	The dot-dashed curve represents changing $\Delta T$,
	the dashed and solid curves represent changing $N_{it}$
	with two different $T_i=20, 100$ GeV and the dotted line is the
	approximate result of the traditional cut based analysis.
	%linearized approximation. 
	}
        \label{fig:SoverB}
\end{figure}

%\section{Results}
%Before the cuts were there were about 600 signal events in o
%In our test case there were about 600 signal events before the cuts were
%applied, and about 100,000 background events.
The traditional cut based analysis left us with some $\sim 5-15$ signal
events and $\sim 5-10$ background events, i.~e., $S/\sqrt{B}\sim 5$.
On the other hand, the MC based method
often managed to completely remove the background, while still preserving
$\sim 50$ signal events.
There are several ways to improve the value of $S/\sqrt{B}$ but they all
come at the cost of longer execution time. The different improvements were:
\begin{itemize}
\item	Use high $T_i$ value
\item	Use smaller $\Delta T$ for each iteration
\item	Increase the number of iterations, $N$.
\item	Change more variables than one, before recomputing $S/\sqrt{B}$
\end{itemize}
For most of these improvements, the general behaviour was that
\begin{equation}
	\frac{S}{\sqrt{B}} \sim 5.1\times t^{0.37}
\end{equation}
where $t$ is the time in seconds.
The only exception was in increasing the number of variables,
which was not profitable.
The $S/\sqrt{B}$ is illustrated in Fig.~\ref{fig:SoverB}.
The values have been averaged over ten different optimization runs.
For the dot-dashed curve, the step $\Delta T$ is modified from
$2^4-2^{-10}$ and divided by two each time. The values of $T_i=20$\% and
$N_{it}=2$. We notice that the curve is levelling out asymptotically.
%./ctrlMC.pl -Temp='20' -Delta='10' -Max='2' -Nstat='10' > DeltaM2_2_4-2_-10.out
The solid and the dashed lines both have $\Delta T=10$ and
the number of iterations goes from $N_{it}=2$ to $512$, multiplied by two
each time. Furthermore, $T_i=20$\% for the dashed line and $T_i=100$\% for
the solid one.
% ./ctrlMC.pl -Temp='100' -Delta='10' -Max='10' -Nstat='10' > Nit_T100_2-512.out

%{\it LHC? LEP, other exp?}
%
%{\it Description of OPAL and variables}
%{\it Better with not so many variables}
%
%{\it $T/=1.2$?}
%
%{\it Statistics? Change: T, dT, logdT, nChanges, continue, Nit}
% TODO Figure
% TODO DeltaT not arbitrarily low
% TODO Roof

%\section{Acknowledgements}

I acknowledge support from the Swedish National Graduate School of Space Technology
and I thank F. Sandin for many useful discussions.

\bibliographystyle{elsart-num.bst} % 

\bibliography{bibtex}        % bibtex.bib is the name of our database

\begin{thebibliography}{1}
\expandafter\ifx\csname url\endcsname\relax
  \def\url#1{\texttt{#1}}\fi
\expandafter\ifx\csname urlprefix\endcsname\relax\def\urlprefix{URL }\fi

\bibitem{2000EPJC...14...73O}
{The OPAL Collaboration}, G.~{Abbiendi}, {et al.}, {Search for unstable heavy
  and excited leptons at LEP2}, European Physical Journal C 14 (2000) 73--84.

\bibitem{Elfgren:2002ck}
E.~Elfgren, Heavy and excited leptons in the opal detector?, Master's thesis,
  Universit{\'e} de Montr{\'e}al (2002).

\bibitem{1992NIMPA.317...47A}
J.~{Allison}, et~al., {The detector simulation program for the OPAL experiment
  at LEP}, Nuclear Instruments and Methods in Physics Research A 317 (1992)
  47--74.

\bibitem{Alexander:1991qw}
G.~Alexander, et~al., {Measurement of the $Z^0$ line shape parameters and the
  electroweak couplings of charged leptons}, Z. Phys. C52 (1991) 175--208.

\bibitem{EXOTIC}
R.~Tafirout, G.~Azuelos, {EXOTIC - A heavy fermion and excited fermion Monte
  Carlo generator for $e^+e^-$ physics}, Computer Physics Communications 126
  (2000) 244--260.

\end{thebibliography}

\end{document}